\documentclass[%
times, twocolumn,
superscriptaddress,
 amsmath,amssymb,
 aps,
]{revtex4-1}

\usepackage[english]{babel}
\usepackage{tikz,xr-hyper}
\usepackage[utf8]{inputenc}

\setcitestyle{sort&compress,numbers}
\usepackage{amsthm, color, mathtools, amsmath, amssymb, setspace,bbm, tensor, subfigure, mathtools, placeins, graphicx, wrapfig,float, verbatim}
\usepackage{graphicx}
\usepackage[unicode=true, breaklinks=false, pdfborder={0 0 1}, backref=false, colorlinks=true, linkcolor=blue, citecolor=blue]{hyperref}
\usepackage{verbatim}
\usepackage{cancel,bm}
\usepackage{subfigure}

\let\oldsqrt\sqrt
\def\sqrt{\mathpalette\DHLhksqrt}
\def\DHLhksqrt#1#2{%
\setbox0=\hbox{$#1\oldsqrt{#2\,}$}\dimen0=\ht0
\advance\dimen0-0.2\ht0
\setbox2=\hbox{\vrule height\ht0 depth -\dimen0}%
{\box0\lower0.4pt\box2}}

\newcommand{\tr}{\operatorname{tr}}

\newcommand{\smn}{s_{-}}
\newcommand{\spl}{s_{+}}
\newcommand{\spm}{s_{\pm}}

\newcommand{\Bcal}{\mathcal{B}}
\newcommand{\Ecal}{\mathcal{E}}

\newcommand{\Hcal}{\mathcal{H}}
\newcommand{\Ical}{\mathcal{I}}

\newcommand{\Tcal}{\mathcal{T}}
\newcommand{\Ucal}{\mathcal{U}}

\newcommand{\Scal}{\mathcal{S}}

\newcommand{\Jcal}{\mathcal{J}}
\newcommand{\Rcal}{\mathcal{R}}

\newcommand{\Pprob}{\mathbb{P}}

\newcommand{\ident}{\mathbbm{1}}

\newcommand{\iu}{{i\mkern1mu}}

\def\bra#1{\mathinner{\langle{#1}|}}

\def\ket#1{\mathinner{|{#1}\rangle}}

\newcommand*\xbar[1]{%
   \hbox{%
     \vbox{%
       \hrule height 0.5pt 
       \kern0.5ex
       \hbox{%
         \kern-0.2em
         \ensuremath{#1}%
         \kern-0.0em
       }%
     }%
   }%
}

\def\BraVert{\egroup\,\mid\,\bgroup}

\def\ketbra#1#2{\ket{#1\vphantom{#2}}\!\bra{#2\vphantom{#1}}}

\def\bra#1{\mathinner{\langle{#1}|}}

\def\ket#1{\mathinner{|{#1}\rangle}}

\usepackage{bbm, braket}
\usepackage[mathscr]{euscript}
\allowdisplaybreaks
\newtheorem*{theorem*}{Theorem}

\newtheorem{lemma}{Lemma}

\newtheorem{definition}{Definition}

\begin{document}

\title{Completely positive divisibility does not mean
Markovianity}

\author{Simon Milz}
\email{simon.milz@monash.edu} 
\affiliation{School of Physics and Astronomy, Monash University, Clayton, Victoria 3800, Australia}
    
\author{M. S. Kim}
\affiliation{QOLS, Blackett Laboratory, Imperial College London, London SW7 2AZ, United Kingdom}
\affiliation{Korea Institute for Advanced Study, 02455, Seoul, Korea}

\author{Felix A. Pollock}
\affiliation{School of Physics and Astronomy, Monash University, Clayton, Victoria 3800, Australia}

\author{Kavan Modi}
\affiliation{School of Physics and Astronomy, Monash University, Clayton, Victoria 3800, Australia}

\begin{abstract}
In the classical domain, it is well-known that divisibility does not imply that a stochastic process is Markovian. However, for quantum processes, divisibility is often considered to be synonymous with Markovianity. We show that completely positive (CP) divisible quantum processes can still involve non-Markovian temporal correlations, that we then fully classify using the recently developed process tensor formalism, which generalizes the theory of stochastic processes to the quantum domain. 
\end{abstract}
\date{\today}
\maketitle

No system is fully isolated from its surroundings. This is especially true for quantum processes, where along with the surrounding environment, the act of observation can disturb the system~\cite{modiosid}. The field of open system dynamics attempts to develop methods that describe the dynamics of systems, quantum and classical, away from isolation~\cite{breuer_theory_2007}. These tools become crucial in analyzing a whole host of problems, from strong coupling thermodynamics~\cite{arXiv:1810.00698} to error correction in quantum technologies~\cite{13preskill}. An important consideration for describing open dynamics is the size and length of memory that the surroundings possess about the system's past~\cite{arXiv:1805.11341, arXiv:1810.10809}. In general the future states of the system depend non-trivially on its own past, leading to complex joint measurement statistics in time~\cite{thao, PhysRevLett.111.020403}. A process where the environment has \textit{no} memory is called Markovian, and the complexity of describing such dynamics scales only as the Hilbert space dimension of the system~\cite{pollock_operational_2018}, while the complexity of a non-Markovian process can scale exponentially in the number of times considered~\cite{pollock_non-markovian_2018, costashrapnel2016}.

Markovianity plays an important role in fields ranging far beyond the physical sciences. This is both for the fact that many processes in nature are approximated sufficiently well by memoryless dynamics, and the computational and simulation intricacy that arises once memory effects are taken into account~\cite{crutchfield_inferring_1989, breuer_theory_2007}. As experimental control over complex quantum systems becomes increasingly sophisticated, the ability to directly determine whether a Markovian description is applicable is becoming ever more important~\cite{Gessner:2014kl, fanchini2014, josh, PhysRevLett.114.090402, winick_phenomenological_2019}. Consequently, in recent years, a large body of work dedicated to the characterization of temporal correlations in quantum systems has arisen~\cite{caruso_quantum_2014,rivas_quantum_2014,de_vega_dynamics_2017, li_concepts_2018}. However, strictly testing for the presence of memory effects is an intractable task in general~\cite{pollock_non-markovian_2018}, and a zoo of non-Markovianity witnesses has emerged in the past two decades~\cite{PhysRevLett.101.150402, PhysRevLett.103.210401, PhysRevLett.105.050403, PhysRevA.82.042103, hou2011, PhysRevA.83.022109, rajagopal,  mazzola2012dynamical, rodriguez2012unification, PhysRevA.86.044101, bylicka2014, sabri, bylicka2016, zhihe2017, chruscinski2018}.

The concept that underpins \textit{all} of these witnesses is completely positive (\textbf{CP}) divisibility, a frequently used proxy for Markovianity. While experimentally accessible~\cite{liu_experimental_2011,tang_measuring_2012, bernardes_experimental_2015, jin_all-optical_2015}, this criterion lacks a clear, quantifiable link to Markovianity, which casts its implications for potential memory effects, and the interpretation of all the memory witnesses derived thereof, into doubt.  In this Letter, we first demonstrate that, \textit{a priori}, there are inequivalent experimental definitions of CP divisibility. After clearing up these ambiguities, we close the fundamental gap in the understanding of CP divisibility and comprehensively derive its quantitative relationship to Markovianity. While the difference between CP divisibility and Markovianity has been pointed out before~\cite{accardi_quantum_1982, pollock_operational_2018}, our results yield both a quantifiable delineation between them, as well as a comprehensive characterization of the temporal correlations CP divisibility is sensitive to. This, in turn, provides a meaningful way forward for experimentalists looking to definitively characterize noise in their devices by means of memory witnesses. To motivate the relation of Markovianity and divisibility we first briefly review them in the context of classical processes. 

\textbf{Markovianity and divisibility---} Mathematically, a classical process is called Markovian if the current state conditionally only depends on the last one, and not the whole history:
\begin{gather}\label{eq:clMarkov}
\!\Pprob(x_n,t_n|x_{n\!-\!1},t_{n\!-\!1};\! \dots \!;x_0,t_0) \!=\! \Pprob(x_n,t_n|x_{n\!-\!1},t_{n\!-\!1}).
\end{gather}
A generalization of this condition to quantum theory has recently been achieved~\cite{pollock_operational_2018}.

From Eq.~\eqref{eq:clMarkov} it is clear that Markovianity is a statement about \textit{multi-time} correlations and checking for it requires an exponentially large set of conditions to be satisfied. A simpler criterion that follows from Markovianity, but is not sufficient to define it, is \emph{divisibility}. This requires the conditional probabilities, for any \emph{three} times $t > s >r$, to factorize according to the Chapman-Kolmogorov equation: $\Pprob(x,t|z, r) = \sum_y \Pprob(x,t|y, s) \Pprob(y,s|z, r)$ for all states $x,y,z$, where each $\Pprob$ is a probability distribution. The natural quantum generalization of a conditional probability distribution (with a single argument) is a completely positive map, \textit{i.e.}, one which preserves the positivity of even correlated density operators~\cite{SudarshanMatthewsRau61}, and the Chapman-Kolmogorov equation generalizes to the condition for CP divisibility~\footnote{We will not consider the notion of P~divisibility, where intermediate maps are only required to be positive.}:
\begin{definition}[CP divisibility]
A quantum dynamical process of a system on an interval $[0,\Tcal]$ is CP divisible if $(i)$ the dynamical map from $r$ to $t$ acting on the system of interest can be broken up at $s$ such that
\begin{gather}\label{eq:divdef}
    \Phi_{t:r} = \Phi_{t:s} \circ \Phi_{s:r} 
\quad \forall \quad \Tcal \ge t \ge s \ge r \ge 0,
\end{gather}
and $(ii)$ each map $\Phi_{x:y}$ is completely positive.
\end{definition}

Intuitively, the connection between this definition and Markovianity is that CP-maps describe dynamics without initial system-environment correlations~\cite{breuer_theory_2007}, and Eq.~\eqref{eq:divdef} suggests that the dynamics between intermediate times are independent of the past. Together, these properties could be taken to imply the absence of memory. However, while mathematically well-defined~\cite{wolf_dividing_2008}, \textit{a priori}, neither the operational meaning of the family of maps $\{\Phi_{t:s}\}$, nor its relation to prevalent memory
effects are clear. That is, in an experimental setting, what exact \textit{quantum process tomography} procedure~\cite{nielsen_quantum_2000} is required to determine CP divisibility, and what exactly does this property imply?

There are (at least) two non-equivalent ways to address these questions. In what follows, we first motivate and define two types of CP divisibility and show their non-equivalence.  We then give a full characterization of the non-Markovian temporal correlations that may hide in a divisible process, thus providing a clear connection and delineation between Markovianity and CP divisibility. Throughout this Letter, we will only consider systems with finite Hilbert-space dimension $d$. 

\begin{figure}
    \centering
\subfigure[~Constructing $\Lambda_{s:0}$]
{\includegraphics[scale=.83]{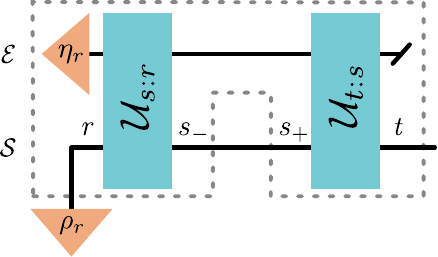}
\label{fig::Dilation1}}
\quad
\subfigure[~Constructing $\Lambda_{t:s}$]
{\includegraphics[scale=.83]{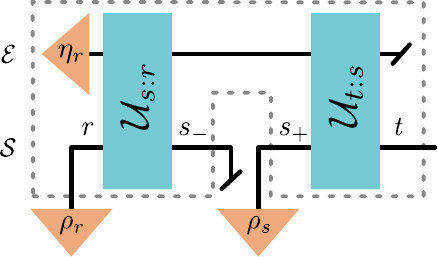}
\label{fig::Dilation2}}
    \caption{\textbf{Circuits for checking iCP and oCP divisibility.} (\textbf{a}) To construct the maps $\Lambda_{s:0}$ and $\Lambda_{t:0}$ we may set $r=0$ and measure the system at $s$ or $t$ respectively. (\textbf{b}) To construct $\Lambda_{t:s}$, the system is discarded at $\smn$ and a fresh state is fed in at $\spl$. The dotted line encapsulates the three-time process tensor $T_{t:s:r}$.}
    \label{fig::Dilations}
\end{figure}

\textbf{CP divisibility by inversion.---}
Consider an experimental setup where one is allowed to prepare any desired state at the initial time $r=0$ and perform measurements on the system at any later time $s$. Within these experimental constraints, using the standard method of quantum process tomography, one can construct a family of maps $\lambda_0 := \{\Lambda_{s:0}\}$ that describe the dynamics from time $r=0$ to time $s$, see Fig.~\ref{fig::Dilation1}. Assuming that all the maps of this family are invertible, we obtain the following definition, which is the one that  most frequently appears in the literature~\cite{PhysRevLett.101.150402,breuer_foundations_2012, BreuerEA2016}:
\begin{definition}[iCP divisibility]
A process is CP divisible by inversion (iCP-divisible) if for any two maps $\Lambda_{s:0}, \Lambda_{t:0} \in \lambda_0$ with $\Tcal\geq t>s\geq 0$ the map
\begin{gather}\label{eq:defindiv}
    \Phi_{t:s} := \Lambda_{t:0} \circ \Lambda_{s:0}^{-1}
\end{gather}
is completely positive.
\end{definition}
Here, we choose a convention where experimentally accessible maps are denoted by $\Lambda$. Notably, if all elements of $\lambda_0$ are invertible then each $\Phi_{t:s}$ constructed according to Eq.~\eqref{eq:defindiv} is well-defined, and can be obtained computationally from $\lambda_0$. 

\textbf{Operational divisibility.---} While iCP divisibility is well-defined, it leaves the operational meaning of the inferred maps $\Phi_{t:s}$ open~\footnote{One possible operational interpretation of iCP-divisible \textit{dynamics} -- but \textit{not} the maps $\Phi_{t:s}$ themselves or the concept's relation to Markovianity -- has been considered elsewhere~\cite{PhysRevLett.117.050403}, but is unrelated to this work.}; particularly, these maps do not necessarily relate to anything that could actually be measured at intermediate times. Additionally, due to their non-operational definition, it is not possible to straightforwardly characterize the memory effects that iCP divisibility is blind to. A more operationally motivated definition of CP divisibility based on experimentally reconstructed maps alone is thus desirable.

To this end, let us consider a scenario where an experimenter can manipulate the system at \textit{any} time $s \, \in [0,\mathcal{T}]$, which we split infinitesimally into $\smn$ and $\spl$ as shown in  Fig.~\ref{fig::Dilation2}. At time $\smn$ the system is discarded and, at $\spl$, replaced with a fresh one in state $\rho_s$. Subsequently, the experimenter measures the system at time $t$. With this procedure they can experimentally reconstruct maps $\lambda := \{\Lambda_{t:s}\}$ as
\begin{gather}
\label{eqn::OpCP}
    \Lambda_{t:s}[\rho_{s}] = \tr_\Ecal\left[\Ucal_{t:s} \left( \rho_{s} \otimes \eta_{s} \right) \right] ,
\end{gather}
where $\eta_{s}$ is the reduced state of the environment at time $s$ and $\Ucal_{t:s}(x_s):= U_{t:s} \, x_s \, U_{t:s}^\dag = x_t$ is the unitary system-environment map. We can thus define oCP divisibility:
\begin{definition}[oCP divisibility]
\label{def:oCP}
A process is operationally CP-divisible (oCP-divisible), if for any $\Tcal\geq t>s>r\geq 0$
\begin{gather}\label{eq:defopdiv}
\Lambda_{t:r} = \Lambda_{t:s} \circ \Lambda_{s:r}
\end{gather}
holds, where the maps above belong to set $\lambda$ and are defined in Eq.~\eqref{eqn::OpCP}.
\end{definition}
Importantly, complete positivity of the respective maps is guaranteed, as system-environment correlations are discarded for their reconstruction. Formally, Eq.~\eqref{eq:defopdiv} resembles Eq.~\eqref{eq:divdef}, with the important distinction that here each map has a clear operational meaning.

Still, there is a level of ambiguity in the reconstruction procedure of the maps $\Lambda_{t:s}$. In principle, they could depend on preparations at any previous time $r$, which would imply non-Markovianity~\cite{pollock2018t}. However, if there are at least two different states $\rho_r$ and $\widetilde{\rho}_r$, such that the corresponding maps $\Lambda_{t:s}$ and $\widetilde{\Lambda}_{t:s}$ differ, then oCP divisibility is not uniquely defined. Thus, Def.~\ref{def:oCP} implicitly requires that the intermediate maps are independent of any earlier state preparations. This independence constitutes a non-signalling condition~\cite{piani_properties_2006, chiribella_perfect_2012, kofler_condition_2013}, as we discuss formally in App.~\ref{app:Lem}.

Importantly, this non-signalling requirement is a \textit{conditional} one; for oCP-divisible dynamics, there is no signalling from $r$ to $t$ given that the system state was discarded at $\smn$. Equivalently, this condition can be thought of as follows: consider an experiment where one part of a correlated state $\rho_{rr'}$ is fed into the process at time $r$. At time $\smn$ the system is discarded and a fresh state prepared at $\spl$, and the experimenter looks for correlation in the resulting state $\rho_{tr'}$ at time $t$. If $\rho_{tr'} \ne \rho_{t} \otimes \rho_{r'}$ then we have conditional signalling from $r$ to $t$.

Conditional non-signalling is, for example, satisfied if the system interacts only once, between any two times, with a part of the environment that is discarded afterwards. However, while conditional non-signalling is necessary for oCP divisibility, it is not sufficient, see App.~\ref{app:notsuff} and also~\cite{PhysRevA.83.052128}.

The absence of signalling is reminiscent of the concept of \emph{no information back-flow} attributed to CP-divisible processes~\cite{PhysRevLett.103.210401}. Here, however, in contrast to the increase of trace distance between trajectories, signalling is a genuine multi-time statement. Now, before further discussing their relationship to Markovianity, we show that iCP and oCP divisibility do not coincide.

\textbf{oCP divisibility $\ne$ iCP divisibility.---} Despite their superficial resemblance, iCP and oCP divisibility differ in their experimental reconstruction and the meaning of the maps they comprise. Consequently, the relationship between them is \textit{a priori} unclear. First, note that iCP divisibility is only defined if all elements of the set $\lambda_0$ are invertible. This limitation does not apply to oCP divisibility. Focussing on the invertible case, we find that oCP divisibility implies iCP divisibility by direct application of Eq.~\eqref{eq:defopdiv}.

To see that the converse does not hold, we construct an iCP-divisible dynamics that is conditionally signalling, and thus not oCP-divisible. Consider the two circuits in Fig.~\ref{fig::Dilations}, where both the system and the environment are qubits, and let the initial environment state be maximally mixed, i.e., $\eta_r = \mathbbm{1}_2/2$. The system-environment dynamics is given by the partial swap  $U_{s:r} = \mathrm{exp}(-\iu \omega \mathbf{S} u) =  \cos(\omega u) \mathbbm{1}_4 - i \sin(\omega u) \mathbf{S}$, where $\mathbf{S}\ket{ij}=\ket{ji}$, and $u := s-r$. We show in App.~\ref{app::oCP_iCP} that the resulting dynamics on the system is iCP-divisible for $\omega t\leq \frac{\pi}{2}$. On the other hand, if we discard the state of the system at $\smn$ and insert a fresh state at $\spl$ we will find that the state at $t$ depends on $\rho_r$ due to the partial swap. In other words we have signalling, and therefore the process is not oCP-divisible.

\begin{figure}
    \centering
\subfigure[~Constructing $\Lambda_{s:0}$]
{\includegraphics[scale=1.4]{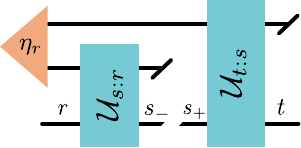}
\label{fig::ExoCP}}
\quad \
\subfigure[~CP Divisibility and Markovianity]
{\includegraphics[scale=1.7]{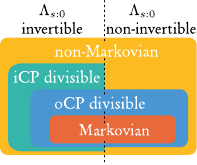}
\label{fig::Venn}}
\caption{\textbf{Divisibility and Markovianity.} (\textbf{a}) The system interacts with one part of the correlated environment state leading to a non-Markovian oCP-divisible process. (\textbf{b}) The hierarchy of sets of processes with varying degrees of temporal correlations.}
\end{figure}

Operationally CP-divisible dynamics form a strict subset of iCP-divisible ones, see Fig.~\ref{fig::Venn}. While the operational requirement is harder to check experimentally, it has a threefold advantage: first the involved maps have a clear-cut operational meaning, and the property of oCP divisibility ties in effortlessly with frameworks tailored for the discussion of non-Markovian quantum processes. Second, the definition of oCP divisibility does not rely on the invertibility of $\Lambda_{s:0}$ and thus has wider applicability. Lastly, oCP divisibility breaks down for a larger class of memory effects than iCP divisibility, and consequently outperforms it as a witness of non-Markovianity.


\textbf{CP divisibility $\ne$ Markovianity.---}
Even though oCP divisibility is a stricter requirement than iCP divisibility, it does not enforce Markovianity; we show this by means of a discrete time example. For an \textit{ante litteram} continuous example of non-Markovian oCP-divisible dynamics, see~\cite{accardi_quantum_1982, arenz_distinguishing_2015}. We take inspiration from collision models~\cite{kim_phase-sensitive_1995, ziman_diluting_2002, scarani_thermalizing_2002, PhysRevA.96.032111} with correlated environment states~\cite{Rybar2012, Bernardes2014}: Let the environment at $r=0$ be in a correlated bipartite state that is uncorrelated with the system. The dynamics $\Ucal_{y:x}$ between any two (of a set of three) times is such that the system only interacts with one part of the environment (denoted by $x$) that is discarded afterwards, see Fig.~\ref{fig::ExoCP}. This scenario satisfies the necessary non-signalling condition. Now, if we choose the unitaries $U_{y:x}$ to be the swap operator $\mathbf{S}_{\Scal x}$ between the system and part $x$ of the environment, then we have $\Lambda_{t:r} = \Lambda_{t:s} \circ \Lambda_{s:r}$, and the dynamics is oCP-divisible.

However, the process is non-Markovian; suppose the experimenter \emph{stores} the system state at time $\smn$, and inserts a fresh state at $\spl$. The dynamics is allowed to continue to $t$ and that state too is stored. The joint state $\rho_{st}$ will be correlated even though the states inserted into the process, at times $r$ and $\spl$, were independent. In particular, for the above case the resulting state $\rho_{st}$ is exactly the correlated initial state of the environment. The experimenter could thus detect memory effects, although the dynamics is oCP-divisible~\cite{vacchini_markovianity_2011}.

Nonetheless, an oCP-divisible process can be seen as one that is Markovian \textit{on average}: Consider a multi-time process where an experimenter measures the system at each time, before independently preparing it in a new state; oCP divisibility implies that, if all past measurement outcomes are forgotten or averaged over then the future statistics only depend on the current preparation. A quantum Markov process, in contrast, requires that the future statistics only depend on the current preparation for \textit{any} sequence of measurement outcomes~\cite{pollock_operational_2018, pollock_operational_2018, oreshkov_causal_2016, costashrapnel2016, sakuldee_non-markovian_2018, giarmatzi_witnessing_2018}.  We now fully characterize the temporal correlations that can persist in oCP-divisible dynamics, thus providing a quantifiable connection between Markovianity and the the majority of witnesses of non-Markovianity employed in the literature.

\textbf{Correlations in divisible processes.---} The four classes of processes illustrated in Fig.~\ref{fig::Venn} also have analogues in the classical domain. A classical stochastic process is described by a joint distribution 
\begin{gather}\label{eq:prob}
\Pprob(x_n,t_n; \ldots ;x_0,t_0),
\end{gather}
over the state of the system at different times, satisfying Kolmogorov conditions~\cite{kolmogorov_foundations_1956, smirne_coherence_2019}. To check if a given process is Markovian necessitates checking all conditional probabilities given in Eq.~\eqref{eq:clMarkov}, which requires the full distribution of Eq.~\eqref{eq:prob}. However, to infer the divisibility of a process, by inversion or operationally, requires only the bipartite marginal distributions of Eq.~\eqref{eq:prob}: $\{\Pprob(x_s,t_s, x_0,t_0)\}_{s=1}^{n}$ and  $\{\Pprob(x_s,t_s, x_r,t_r)\}_{s>r \geq 0}^{n}$ respectively. Thus we have the same hierarchy as in  Fig.~\ref{fig::Venn} for temporal correlations in classical processes.

The quantum generalization of Eq.~\eqref{eq:prob} is a multipartite positive operator $T_{n:\ldots:1:0}$~\cite{Lindblad1979, modi-scirep, chiribella_quantum_2008, chiribella_theoretical_2009, pollock_non-markovian_2018, costashrapnel2016}, called the \textit{process tensor} in the field of open quantum system dynamics~\cite{pollock_non-markovian_2018}, which satisfies \textit{generalized}  Kolmogorov conditions~\cite{accardi_quantum_1982, arXiv:1712.02589}. Analogous to the classical case, the process tensor captures all temporal correlations in quantum processes, including across multiple time steps, in our case three. The probability of observing a sequence of events $\{x_r, x_s, x_t\}$, can be computed by contracting it with generalized measurement operators $M_x$:
\begin{gather}
\label{eqn::ProcTens}
\Pprob(x_t,x_s,x_r|\Jcal_t,\Jcal_s,\Jcal_r) \!= \! \tr[(\! M_{x_t} \! \otimes \! M_{x_s} \! \otimes \! M_{x_r} \!)T_{t:s:r}],
\end{gather}
which constitutes a generalization of the Born rule to processes in time~\cite{chiribella_memory_2008, shrapnel_updating_2017}; $\Jcal$ denotes an instrument~\cite{holevo_1982}, which is a collection of conditional transformations (CP maps) $\{M_{x_s}\}$ that update the system after a particular event is observed; these generalize the concept of positive operator valued measure (POVM). Without loss of generality, each element of Eq.~\eqref{eqn::ProcTens} is expressed in terms of Choi states~\cite{Choi1975, jamiolkowski_linear_1972, chiribella_theoretical_2009, milz_introduction_2017}.

Mathematically, the process tensor $T:=T_{t:s:r}$ is an operator on Hilbert spaces $\Hcal_r \otimes \Hcal_{\smn} \otimes \Hcal_{\spl} \otimes \Hcal_t$. For both panels in Fig.~\ref{fig::Dilations}, the process tensor (the object within the dotted lines) is exactly the same. The difference between them lies entirely in the instrument at $s$. The instrument at $r$ is a preparation with one element $M_{x_r} = \rho_r$, and the instrument at $t$ is a measurement $\{M_{x_t}= \Pi_{x_t}\}$, where the latter are POVM elements. The instrument at $s$ for Fig.~\ref{fig::Dilation1} implements the identity channel, which has Choi state $M_{x_s} = \varphi^+_{\spm}$, where $\varphi^+_{\spm} := \sum_{jk}^d \ket{jj}\bra{kk}$ and $s_\pm := s_-s_+$. For Fig.~\ref{fig::Dilation2} the instrument at $s$ also has a single element: $M_{x_s} = \mathbbm{1} \otimes \rho_s$, which denotes the trace at $\smn$ followed by a preparation of  $\rho_s$. We review the details of the process tensor formalism in App.~\ref{app::ProcDet} and only include important details here.

Using Eq.~\eqref{eqn::ProcTens} and the details of the instruments, we recover the maps in Eq.~\eqref{eq:defopdiv} from the process tensor. Let $L_{x:y}$ denote the Choi state of $\Lambda_{x:y}$. For an oCP-divisible process, we can show that $L_{t:r} = \tr_{\spm}(\varphi_{\spm}^+ T)$, while $L_{s:r} = \tr_{\spl t}[T]/d$ and  $L_{t:s} = \tr_{r\smn}[T]/d$. With this, we can rephrase oCP divisibility as 
\begin{gather}
\label{eqn::oCPChoi}
  \tr_{s}(\varphi_{\spm}^+ T) = \frac{1}{d^2} \tr_{\spm}\left[\tr_{r\smn}(T) \varphi_{\spm}^+ \tr_{\spl t}(T)\right].
\end{gather}
A detailed derivation of above statements is given in Apps.~\ref{app::CondProc} and~\ref{app:last}.

On the other hand, a quantum process is Markovian iff the Choi state of the corresponding process tensor has the form $T^{\mathrm{Markov}} = L_{t:s} \otimes L_{s:r}$~\cite{pollock_operational_2018, costashrapnel2016, oreshkov_causal_2016, giarmatzi_witnessing_2018}; any deviation from this product form implies detectable non-Markovian correlations. Since Eq.~\eqref{eqn::oCPChoi} does not force $T$ to be of Markov form, oCP-divisible processes are not necessarily memoryless. Specifically, representing $T = L_{t:s} \otimes L_{s:r} + \chi_{tsr}$, where the matrix $\chi$ contains all tripartite non-Markovian correlations and satisfies $\tr_{\smn r}[\chi_{tsr}] = \tr_{t\spl}[\chi_{tsr}] = 0$, we see that Eq.~\eqref{eqn::oCPChoi} implies $\tr_{\spm}(\varphi^+_{\spm} \chi_{tsr}) = 0$, which provides a full classification of non-Markovian temporal correlations that can be present in oCP-divisible processes.

\textbf{Conclusions.---} In this Letter, we have provided an operationally motivated definition of CP divisibility that is stricter than the frequently used one relying on the invertibility of $\Lambda_{s:0}$. We showed that oCP divisibility is closely connected to non-signalling conditions and implies the absence of information flow from the environment to the system. Additionally, we have demonstrated that oCP divisibility can be interpreted as Markovianity on average, yet oCP divisible processes can still display non-trivial memory effects, which we have fully characterized. These results lay the foundation for a quantitative interpretation of all studies of memory effects that are based upon CP divisibility or witnesses derived thereof.

Near-term quantum technologies will require effective methods for detecting and addressing non-Markovian noise~\cite{nisq}. We have shed light on divisibility from an operational point of view, which helps us to identify the classes of temporal correlations that may evade regularly used checks for non-Markovianity. However, there are trade-offs between uncovering temporal correlations and the requisite number of experiments that must be performed. Our results enable experimentalists to make informed decisions about investing resources in classifying the non-Markovian noise at hand.

\begin{acknowledgments}
MSK is supported by Samsung Global Research Outreach (GRO) project, the Korea Institute of Science and Technology (KIST) Institutional Program (2E26680-18-P025) and the Royal Society. SM is supported by the Monash Graduate Scholarship, the Monash International Postgraduate Research Scholarship and the J. L. William Scholarship.  KM is supported through Australian Research Council Future Fellowship FT160100073.
\end{acknowledgments}

\FloatBarrier

\bibliographystyle{apsrev4-1}
\bibliography{references}

\appendix

\section{Alternative proof for conditional non-signalling}
\label{app:Lem}
In the main text, we argue that oCP divisibility requires conditional non-signalling between times $r$ and $t$ as, otherwise, it would not be uniquely defined. We can state this more formally:
\begin{lemma}
\label{lem::Non-signal}
oCP divisibility requires that there is no signalling from $r$ to $t$ when the system is discarded and freshly prepared at $\spl$, as depicted in Fig.~\ref{fig::Dilation2}.
\end{lemma}
Another way to prove this lemma  than the one alluded to in the main text is to show that the map $\Lambda_{t:s}$ cannot be CP if there is conditional signalling. To see this, suppose the state at $r$ is chosen to be $\rho_r^{(k)}$ and at $s$ the system is discarded and freshly prepared in a fixed state $\rho_s$. Then the dynamics is allowed to continue to $t$. We will find the state at time $t$ to be $\rho_t^{(k)}$. If the dynamics is conditionally signalling, there is at least one state $\rho_s$ that we can prepare at time $s$ (and that we assume the experimenter prepared), such that the final state at time $t$ depends on the choice of $\rho_r^{(k)}$, thus there exists a dependence of $\rho_t^{(k)}$ on $k$. This means that the dynamics from $s$ to $t$ cannot be CP, because such dynamics always contract information. Since we always prepare the same state at $s$ for CP dynamics we should always receive a single state at $t$, which implies a conditional non-signalling condition in order for $\Lambda_{t:s}$ to be CP.

\begin{figure}
    \centering
    \includegraphics[width=0.85\linewidth]{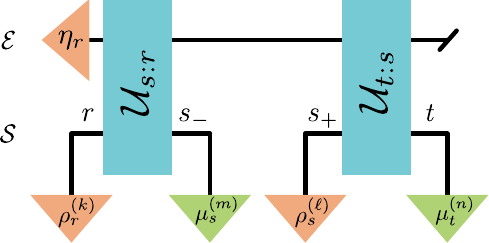}
    \caption{Experimental setup to probe non-signalling conditions. For $x=1$ (see text), the system is measured at times $\smn$ and $t$ with corresponding POVM elements $\mu_s^{(m)}$ and $\mu^{(n)}_t$, and fresh states are fed into the process at times $r$ and $\spl$. If $x=1$, nothing is done at time $s$, \textit{i.e.}, the system wire between $\Ucal_{s:r}$ and $\Ucal_{t:s}$ is uninterrupted.}
    \label{fig::x_equals_0}
\end{figure}

Put more rigorously, conditional non-signalling can be phrased as follows: We fix a tomographically complete set of input states $\{\rho^{(k)}_r\}$ for time $r$ and a tomographically complete POVM, with operators $\{\mu^{(n)}_t\}$, at time $t$. At time $\smn$, we have two choices denoted by the variable $x\in\{0,1\}$: for $x=0$, we `do nothing' at time $s$, \textit{i.e.}, we apply the identity POVM followed by the null preparation. For $x=1$ we apply a fixed tomographically complete POVM, with operators $\{\mu^{(m)}_s\}$ followed by a preparation of a tomographically complete set of input states $\{\rho^{(\ell)}_s\}$ (see Fig.~\ref{fig::x_equals_0} for a graphical representation). 
In the latter case (\textit{i.e.}, $x=1$), the environment state at time $\spl$ is -- up to normalization -- given by
\begin{gather}
    \eta_s^{(mk)} = \tr_\Scal\left\{\mu_s^{(m)} \Ucal_{s:r}[\rho^{(k)}_r \otimes \eta_r]\right\}\, ,
\end{gather}
where $\tr_\Scal$ denotes a trace over the system degrees of freedom. With this, the probability to measure outcomes $m$ and $n$ at times $\smn$ and $t$ given the preparations at times $r$ and $\spl$ can be computed via
\begin{gather}
\Pprob(nm|\ell k,x=1) =
\tr\{\mu_t^{(n)} \Ucal_{t:s} [\rho_s^{(\ell)} \otimes \eta_s^{(mk)}] \} 
\end{gather}
Now, the underlying process would be non-signalling from the preparation and the measurement at times $r$ and $\smn$ to the time $t$ for this experimental situation, if it satisfied $\Pprob(nm|\ell k,x=1)= \Pprob(nm'|\ell k',x=1)$ for all $k,k',m,m'$. It is \textit{conditionally} non-signalling, if it is non-signalling from time $r$ to time $t$, given that the measurement outcomes at time $\smn$ where discarded, \textit{i.e.}, if it satisfies
\begin{gather}
    \Pprob(n|\ell k,x=1) = \Pprob(n|\ell k',x=1)\, ,
\end{gather}
where we have set $\Pprob(n|\ell k) = \sum_m \Pprob(nm|\ell k)$ and $\Pprob(n|\ell k') = \sum_{m'} \Pprob(nm'|\ell k')$. The respective maps involved in the definition of oCP divisibility are then given by   $\Pprob(n|k,x=0)$ (for $\Lambda_{t:r}$), $\Pprob(m|k,x=1)$ (for $\Lambda_{s:r}$), and $\Pprob(n|\ell,x=1)$ (for $\Lambda_{t:s}$).

\begin{figure}
    \centering
    \includegraphics[width = 0.95\linewidth]{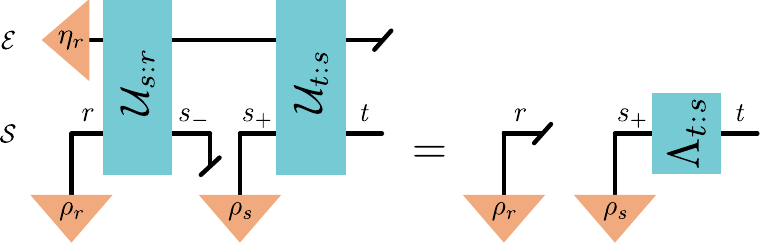}
    \caption{Conditional non signalling. If the dynamics is conditionally non-signalling in the sense of Lem.~\ref{app:Lem}, then system's evolution after discarding it (given here by $\Lambda_{t:s}$ is independent of the input state $\rho_r$ at time $r$). This independence in general does not hold if the system is measured at time $\smn$.}
    \label{fig::Cond_non_signalling_App}
\end{figure}

As mentioned above, intuitively, the requirement of conditional non-signalling implies that if the state of the system is traced out at time $\smn$ then the dynamics from time $\spl$ to time $t$ is independent of the state of the system that is fed into the dynamics at time $r$. A graphical representation of this condition can be found in Fig.~\ref{fig::Cond_non_signalling_App}. While it -- as mentioned in the main text -- corresponds to Markovianity \textit{on average}, it does not amount to the absence of memory. In particular, if the system is measured (in contrast to being discarded) at $\smn$, then the its subsequent dynamics (the map $\Lambda_{t:s}$ in Fig.~\ref{fig::Cond_non_signalling_App}) can in principle depend on the system state at time $r$. Such a dependence would signify the presence of memory effects. 

\section{Conditional non-signalling processes that are not oCP-divisible}
\label{app:notsuff}
As mentioned in the main text, conditional non-signalling is necessary for oCP divisibility to hold, but not sufficient. To see this, consider the following example: Let $\eta_{\Ecal_r\Ecal_s} = \frac{1}{4}\left(\mathbbm{1}_{\Ecal_r\Ecal_s} + \sigma_{\Ecal_r}^{(x)} \otimes \sigma_{\Ecal_s}^{(z)}\right)$ be a correlated two-qubit environment state with $\Ecal_r\Ecal_s:=\Ecal$, where $\{\sigma^{(x)}, \sigma^{(y)}, \sigma^{(z)}\}$ are the Pauli matrices. Initially, i.e., at time $r$, the one-qubit system is uncorrelated with the environment. Let the system-environment dynamics between time $r$ and time $s$ be given by the swap $S_{\Scal \Ecal_r}$. If the system-environment dynamics from $s$ to $t$ only acts non-trivially on $\Scal$ and $\Ecal_s$, then there is no conditional signalling between times $r$ and $s$. Nonetheless, the process is not necessarily CP-divisible. For example, if the unitary evolution between $s$ and $t$ is given by the unitary matrix $U_{t:s} = \frac{1}{\sqrt{3}}\left(\mathbbm{1}_{\Scal\Ecal_s} + \iu\sigma_\Scal^{(y)} \otimes \mathbbm{1}_{\Ecal_s} + \iu\sigma_\Scal^{(x)} \otimes \sigma_{\Ecal_s}^{(z)}  \right)$. With this, the final system state at time $t$, without intervention at time $s$ is $\widetilde{\rho} = \frac{1}{2}\mathbbm{1}_\Scal +\frac{1}{3} \sigma_\Scal^{(y)}$, independently of the input state at time $r$. Consequently, the action of the overall map $\Lambda_{t:r}$  can be written as $\Lambda_{t:r}[\rho] = \tr(\rho)\widetilde{\rho}$. On the other hand, the map $\Lambda_{s:r}$ simply replaces the system state at $r$ with $\tr_{\Ecal_s}\left(\eta_{\Ecal_r\Ecal_s}\right) = \frac{1}{2}\mathbbm{1}_{\Ecal_r}$, which means that, for oCP divisibility to hold, the map $\Lambda_{t:s}$ would have to be of the form $\Lambda_{t:s}[\rho] = \tr(\rho)\widetilde{\rho}$. However, it is easy to check that, for an input state 
\begin{gather}
\rho = \frac{1}{2}(\mathbbm{1}_\Scal + a \sigma_\Scal^{(x)} + b \sigma_\Scal^{(y)} + c \sigma_\Scal^{(z)}),
\end{gather}
at time $s$. The corresponding output state at time $t$ under action of $\Lambda_{t:s}$ is given by
\begin{gather}
 \Lambda_{t:s}[\rho] = \frac{\mathbbm{1}_\Scal}{2} + \frac{a - 2b}{6} \sigma_\Scal^{(x)} + \frac{b}{2} \sigma_\Scal^{(y)} + \frac{2a - c}{6} \sigma_\Scal^{(z)},
\end{gather}
and the process is therefore not oCP-divisible. 

\section{oCP divisibility \texorpdfstring{$\ne$}{} iCP divisibility}
\label{app::oCP_iCP}
Let the system and the environment both be qubits, and let the initial environment state at time $r$ be maximally mixed, i.e., $\eta_r = \mathbbm{1}/2$. In what follows, without loss of generality, we choose $r=0$. The system-environment dynamics is given by a partial swap 
\begin{gather}
    \label{eqn::PartSwap}
    U_{s:0} = \mathrm{exp}(-\iu \omega \mathbf{S} s) =  \cos(\omega u) \mathbbm{1}_4 - i \sin(\omega s) \mathbf{S}
\end{gather} 
where $\mathbf{S}\ket{ij}=\ket{ji}$. For these dynamics, the system state at time $s$ is given by $\rho_s = \cos^2(\omega s)\rho_0 + \sin^2(\omega s)\mathbbm{1}_2/2$, where $\rho_0$ is the system state at time $r=0$, i.e., for all $\omega s\in [0,\pi/2]$, all system states move towards the center of the Bloch ball. Denoting the identity map and the point map, that replaces every state by $\mathbbm{1}/2$, by $\Ical$ and $\Rcal_{\mathbbm{1}}$, respectively, we see that 
\begin{gather}
    \Lambda_{s:0} = \cos^2(\omega s)\Ical + \sin^2(\omega s)\Rcal_{\mathbbm{1}},
\end{gather}
which is invertible for $\omega s \in [0,\pi/2)$. We have 
\begin{align}
\notag
    \Phi_{t:s} &= \Lambda_{t:0} \circ \Lambda_{s:0}^{-1} \\
    &= \frac{\cos^2(\omega t)}{\cos^2(\omega s)}\Ical + \frac{\cos^2(\omega s)-\cos^2(\omega t)}{\cos^2(\omega s)}\Rcal_\mathbbm{1}
\end{align}
which is CP for $t\geq s$ with $\omega t \in [0,\pi/2)$, and, consequently, the dynamics is iCP-divisible in this interval. However, it is not oCP-divisible. Between time $r=0$ and $s$, the environment is partially swapped with the initial state $\rho_0$. Subsequently, after the system state is discarded and freshly prepared at time $s$, between time $s$ and $t$, the system is partially swapped with the environment state, which depends on the state of the system at $r=0$. Consequently, there is conditional signalling from $r$ to $t$, and by Lem.~\ref{lem::Non-signal} the process is not oCP-divisible.

\section{Details of the process tensor formalism}
\label{app::ProcDet}
Here, we give a brief introduction to the process tensor formalism used in the main text. For a more thorough introduction, see, for example, Refs.~\cite{chiribella_theoretical_2009,pollock_non-markovian_2018}. For completeness, we firstly recall the definition of the Choi-Jamio{\l}kowski isomorphism~\cite{Choi1975, jamiolkowski_linear_1972}: Any map $\Gamma: \Bcal(\Hcal_a) \rightarrow \Bcal(\Hcal_b)$ can be mapped isomorphically onto a matrix $G\in \Bcal(\Hcal_b\otimes \Hcal_a)$ by letting it act on one half of an (unnormalized) maximally entangled state $\varphi^+ = \sum_{i,j=1}^d \ketbra{ii}{jj} \in \Bcal(\Hcal_a \otimes \Hcal_a)$, i.e., $G = (\Gamma \otimes \Ical)[\varphi^+]$, where $\Bcal(\Hcal_x)$ is the space of bounded operators on the Hilbert space $\Hcal_x$. 

In classical physics, a process is fully described by a joint probability distribution that allocates the correct probability to all sequences of measurement outcomes at the times of interest. In quantum mechanics, a measurement at time $t_j$ is described by an instrument $\Jcal_j = \{M_{x_j}\}$, a collection of CP maps $M_{x_j} \in \Bcal(\Hcal_{j_+} \otimes \Hcal_{j_-})$ -- represented by their respective Choi states -- that describe the change of the interrogated quantum state upon observation of outcomes $\{x_j\}$~\cite{holevo_1982}. Consequently, a quantum process on times $\{t_n, \dots, t_1\}$ is fully described, once all the probabilities $\Pprob(x_n,\dots,x_1|\Jcal_n, \dots, \Jcal_j)$ for all possible sequences $\{x_n, \dots, x_1\}$ of outcomes for all sequences $\{\Jcal_N,\dots,\Jcal_1\}$ of instruments are known. Due to the probabilistic structure of quantum mechanics, we can compute these probabilities via 
\begin{align}
\label{eqn::Probs}
\notag
&\Pprob(x_n,\dots,x_1|\Jcal_n, \dots, \Jcal_j) \\&\quad= \tr\left[\left(M_{x_n} \otimes \cdots \otimes M_{x_1}\right) T_{n:\dots:1}\right]\, ,
\end{align}
where the positive matrix $T_{n:\dots:1} \in \Bcal(\Hcal_{N_+} \otimes\Hcal_{N_-} \otimes \cdots \otimes \Hcal_{1_-})$ is the \textit{process tensor} of the process, and Eq.~\eqref{eqn::Probs} is the generalization of the Born rule to temporal processes~\cite{shrapnel_updating_2017}. The process tensor then contains all probable multi-time correlations of the process at hand.

Employing this concept to the case of times $\{r,s,t\}$ yields Eq.~\eqref{eqn::ProcTens}, with the physical intuition that a state $\rho_r = M_{x_r} \in \Bcal(\Hcal_{r})$ is prepared at time $r$, interrogated at time $s$, with corresponding CP map $M_{x_s} \in \Bcal(\Hcal_{\spl} \otimes 
\Hcal_{\smn})$, and measured at time $t$, with the outcome corresponding to a POVM element $M_{x_t}\in \Bcal(\Hcal_{t})$.

\section{Conditional non-signalling and marginal channels}
\label{app::CondProc}
Using the formalism reiterated in Sec.~\ref{app::ProcDet}, here, we re-write the requirement of conditional non-signalling in terms of the process tensor formalism. Using the notation of the main text, we see that the process tensor $T_{t:s:r}$ is defined on $\Bcal(\Hcal_t\otimes \Hcal_{\spl}\otimes \Hcal_{\smn}\otimes \Hcal_r)$. Conditional non-signalling means that the final state at time $t$ is independent of the state $\rho_r$ that was prepared at time $r$, if the system was discarded at time $\smn$ and re-prepared at time $\spl$. Under the CJI, the operation of replacing the system with a fresh state $\rho_{\spl}$, corresponds to a matrix $\rho_{\spl} \otimes \ident$. With this, conditional non-signalling can be phrased as
\begin{align}
    &\tr_{r\spl\smn}\left[\left(\rho_r \otimes \mathbbm{1}_{\smn} \otimes \rho_{\spl} \otimes \mathbbm{1}_t\right)T\right] \\ \notag
    &= \tr_{r\spl\smn}\left[\left(\rho_r' \otimes \mathbbm{1}_{\smn} \otimes \rho_{\spl} \otimes \mathbbm{1}_t\right)T\right], \quad \forall \, \rho_r, \, \rho_r', \, \rho_{\spl}
\end{align}
where we have set $T := T_{t:s:r}$. Evidently, this requirement is satisfied iff $\tr_{\smn}T= \mathbbm{1}_r \otimes L_{t:{\spl}}$, with $L_{t:{\spl}}$ the Choi state of the reconstructible channel from time $s$ to time $t$. This requirement is reminiscent of the causality constraints that hold for quantum combs~\cite{chiribella_quantum_2008, chiribella_theoretical_2009}, which follow from non-signalling requirements as well. With this, it is possible to obtain the channel from $s$ to $t$ via $L_{t:\spl} = \tr_{r\smn}[T]/d$; importantly, this tracing procedure would in general only yield the averaged channel between time $s$ and time $t$~\cite{pollock_non-markovian_2018}, but here, due to the conditional non-signalling condition, the averaged channel coincides with the reconstructed channel $L_{t:s}$. The Choi state $L_{t:r}$ of the channel $\Lambda_{t:r}$ from $r$ to $t$ is obtained reconstructively, by `doing nothing' at time $s$. Under the CJI, the do-nothing operation $\Ical$ at time $s$ corresponds to the (unnormalized) maximally entangled state $\varphi^+_{\spl\smn} = \sum_{i,j=1}^d \ketbra{ii}{jj} \in \Bcal(\Hcal_{\spl} \otimes \Hcal_{\smn})$, and as such, $L_{t:r}$ can be obtained from $T$ by contracting it with $\varphi^+_{\spl\smn}$, i.e., 
\begin{gather}
    L_{t:r} = \tr_{\smn\spl}\left[\left(\mathbbm{1}_{rt} \otimes \varphi_{\spl\smn}\right)T\right].
\end{gather}
Finally, the Choi state $L_{t:\spl}$ of $\Lambda_{t:s}$ is obtained by simply tracing out the degrees of freedom of $T$, that belong to the times $\spl$ and $t$~\cite{pollock_non-markovian_2018}:
\begin{gather}
    L_{t:\spl} = \frac{1}{d} \tr_{\spl t}T.
\end{gather}
\section{Derivation of \texorpdfstring{Eq.~\eqref{eqn::oCPChoi}}{}}
\label{app:last}
A dynamics is oCP-divisible, if we have $\Lambda_{t:r} = \Lambda_{t:s} \circ \Lambda_{s:r}$. The Choi states of the CPTP maps $\Lambda_{t:s}$ and $\Lambda_{s:r}$ are given above and in the main text in terms of the process tensor $T$ as 
\begin{gather}
\label{eqn::Lexpr}
L_{\smn:r} = \tr_{\spl t}[T]/d \ \ \text{and} \ \ L_{t:\spl} = \tr_{r\smn}[T]/d
\end{gather}
Now, concatenation of the two maps means, that the output of $\Lambda_{s:r}$ is the input of $\Lambda_{t:s}$, or, equivalently, that an identity map $\Ical$ is `performed' between them at time $s$. As the identity map corresponds to $\varphi^+_{\spl\smn}$, this concatenation $\Lambda_{t:s} \circ \Ical \circ \Lambda_{s:r}$ in terms of their Choi states is expressed as~\cite{chiribella_quantum_2008,chiribella_theoretical_2009} 
\begin{gather}
    \tr_{\spl\smn}[\mathbbm{1}_{r\smn}\otimes L_{t:\spl}\mathbbm{1}_{rt}\otimes \varphi^+_{\spl\smn}\mathbbm{1}_{\spl t}\otimes L_{\smn:r}],
\end{gather}
where the intermediate identity map can be seen as a simple relabelling $B\mapsto C$. Inserting the  expressions in Eq.~\eqref{eqn::Lexpr} for $L_{t:s}$ and $L_{s:r}$ then yields Eq.~\eqref{eqn::oCPChoi}.

\end{document}